\ifx\pdftexversion\undefined
  \documentclass[preprint,dvips]{aastex}
\else
  \documentclass[preprint,pdftex]{aastex}
\fi
\newcommand{\rmmat}[1]{{\hbox{\rm #1}}}
\newcommand{\rmscr}[1]{{\rmmat{\scriptsize #1}}}
\newcommand{\be}{\begin{equation}}
\newcommand{\ee}{\end{equation}}
\newcommand{\bt}{\begin{table} \begin{center}}
\newcommand{\et}{\end{center} \end{table}}
\newcommand{\ba}{\begin{eqnarray}}
\newcommand{\ea}{\end{eqnarray}}
\newcommand{\ie}{{\it i.e.~}}
\newcommand{\eg}{{\it e.g.~}}
\newcommand{\cf}{{\it c.f.~}}
\newcommand{\eqref}[1]{Equation~(\ref{eq:#1})}

\newcommand{\figref}[1]{Figure~\ref{fig:#1}}

\begin{document}
\newcommand{\bfi}{{\bf B}} \newcommand{\efi}{{\bf E}}
\newcommand{\lel}{{\lambda_e^{\!\!\!\!-}}}
\newcommand{\me}{m_e}
\newcommand{\mcs}{{m_e c^2}}
\newcommand{\ho}{{\hat {\bf o}}}
\newcommand{\hm}{{\hat {\bf m}}}
\newcommand{\hx}{{\hat {\bf x}}}
\newcommand{\hy}{{\hat {\bf y}}}
\newcommand{\hz}{{\hat {\bf z}}}
\newcommand{\hr}{{\hat {\bf r}}}
\newcommand{\omv}{\mathbf{\omega}}

\title{The High-Energy Polarization-Limiting Radius of Neutron Star
 Magnetospheres II -- Magnetized Hydrogen Atmospheres}
\author{Jeremy S. Heyl$^{\star}$\footnote{Chandra Postdoctoral
 Fellow, Current Address: Department of Physics and Astronomy, 
University of British Columbia 
6224 Agricultural Road, Vancouver, British Columbia, Canada, V6T 1Z1},
Don Lloyd$^{\star}$, and
Nir J. Shaviv$^{\dag}$}
\affil{$^{\star}$
Harvard College Observatory, MS-51, 
60 Garden Street, Cambridge, Massachusetts 02138, United States}
\affil{$^{\dag}$
Racah Inst.\ of Physics, Hebrew University of Jerusalem, Jerusalem 91904, Israel}

\begin{abstract}
  In the presence of strong magnetic fields, the vacuum becomes a
  birefringent medium. We show that this QED effect couples the
  direction of the polarization of photons leaving the NS surface, to
  the direction of the magnetic field along the ray's path. We analyze
  the consequences that this effect has on aligning the polarization
  vectors to generate large net polarizations, while considering
  thermal radiation originating from a thermal hydrogen
  atmosphere. Counter to previous predictions, we show that the
  thermal radiation should be highly polarized even in the optical.
  When detected, this polarization will be the first demonstration of
  vacuum birefringence. It could be used as a tool to prove the high
  magnetic field nature of AXPs and it could also be used to constrain
  physical NS parameters, such as $R/M$, to which the net polarization
  is sensitive.
\end{abstract}

\section{Introduction}
\label{sec:introduction}

Neutron stars provide an exciting laboratory to understand extreme
realms of physics.  Processes in their interiors probe nuclear theory
in uncharted regimes, and not only is the surrounding spacetime
strongly curved but also it is often pervaded by a strong magnetic
field.  As radiation from the surface of the star traverses this
region, both the curvature of the spacetime and the strong magnetic
field leave traces.  It is possible to unravel these two processes to
verify that the magnetic field of the star does indeed interact with
the outgoing radiation (a prediction of quantumelectrodynamics) and
that the spacetime exterior to the star curves the trajectories of the
rays.  

In several earlier papers \citep[Paper
I:][]{Heyl01polar,Heyl01qed,Shan04} we have examined how the
magnetosphere of a neutron star affects the propagation of polarized
radiation from the surface of the star.  In all of these papers we
assumed a very simple model for the atmospheric emission: the emission
is fully polarized at all frequencies.  In this paper we combine a
realistic treatment of the radiative transfer of the magnetosphere as
in the earlier works with an accurate calculation of the radiative
transfer within in the highly magnetized hydrogen atmosphere of the neutron
star \citep{Lloy03}.  Although several authors have recently focussed
on the emission properties of neutron star atmospheres \citep[for
example][]{2003ApJ...588..962L,2001ApJ...563..276O,2001ApJ...560..384Z},
these techniques have focussed on the case where the magnetic field is
perpendicular to the surface of the star.  The models used here
consider the important, general (and more difficult to calculate) case
where the magnetic points in a arbitrary direction.  Furthermore, we
have used the envelope models of \citet{Heyl97analns} and
\citet{Heyl98numens} to determine the flux at each point on the star.

Although these atmosphere models treat general magnetic geometry, they
lack several features included in other models.  These effects such as
vacuum polarization in the atmosphere itself
\citep{2003PhRvL..91g1101L}, the proton-cyclotron resonance
\citep{2001ApJ...560..384Z,2003ApJ...583..402O} and partial ionization
\cite{2003ApJ...599.1293H} may be important to understand some details
of the emission at certain magnetic field strengths.  However, the
general properties of the polarization spectral energy distribution of
magnetized neutron-star atmospheres is unlikely to depend strongly on
these processes.

Although such mangetized hydrogen atmospheres provide a good
explanation of the data from some neutron stars
\citep{Pern00pulsar,Pern00axp}, some of the more well studied objects
such as  RX~J1856.5-3754 remain inscrutable
\citep{2004ApJ...603..265T}.  \citet{Chan04} has argued
that the hydrogen atmosphere of a magnetar ($B \gtrsim 10^{14}$~G) is
consumed by diffusive nuclear burning within a few thousand years,
leaving a surface layer of helium.  On the other hand, the hydrogen
layer reamins on the surface of more weakly magnetized neutron stars
for up to one hundred times longer \citep{2004ApJ...605..830C}.
Because our focus here is on both magnetars and more weakly magnetized
young neutron stars, we will assume that hydrogen comprises the
atmospheres of these objects.  It is likely that a fully ionized
helium atmosphere exhibits grossly similar emission properties.

In the following sections we describe the calculations in modest
detail (\S~\ref{sec:calculations}) -- the reader is encouraged to look
at the authors' earlier works for further details
\citep[e.g.][]{Heyl01polar,Lloy03}.  A detailed exposition of the
results of these calculations (\S~\ref{sec:results}) follows.  This
includes a comparison with earlier results
(\S~\ref{sec:comp-with-earl}).  In \S~\ref{sec:discussion} we place

our results in the greater context.

\section{Calculations}
\label{sec:calculations}

Calculating the emergent polarized spectra from neutron stars requires
several ingredients.  As we argued in Paper~I, the general
relativistic corrections to the field geometry may be neglected, so we
assume that the field geometry is a centered dipole.  We use the
results of Heyl \& Hernquist (1998, 2001)
\nocite{Heyl97analns,Heyl98numens} to determine the emergent energy
flux over the surface of the star.  We use the results of
\citet{Lloy02} and \citet{Lloy03} to calculate the emergent photon
spectrum at a grid of points on the surface.  Finally, we construct
the image of the neutron star surface by tracing rays from the surface
of the neutron star to the detector.  The final spectrum is obtained
by interpolating the polarized intensities from the grid of atmosphere
models and summing over the calculated image.  In the next
subsections, we will outline each step of this process in further
detail.

\subsection{Atmospheric Spectra}
\label{sec:atmospheric-spectra}

We assume that the total flux at each point $(\phi,\beta)$ of the NS
surface is proportional to 
\begin{equation}
F \propto B^{0.4} \cos^2\psi, 
\label{eq:temp-distribution}
\end{equation}
where
$\psi$ is the angle between the local normal and the direction of the
magnetic field and $B$ is the magnitude of the field. This was shown
to result from the effect that the magnetic field has on heat transfer
throught he NS crust \citep{Heyl97analns,Heyl98numens}.  We use
atmosphere models calculated at $B=10^{12}$~G and $10^{14}$~G.  For
the stronger field, the atmosphere calculation converges much more
quickly for higher fluxes, so we use a model with
$T_\rmscr{p,eff}=10^{6.5}$~K.  At $10^{12}$~G we use
$T_\rmscr{p,eff}=10^{6.5}$~K for comparison with the $10^{14}$~G results
and $T_\rmscr{p,eff}=10^6$~K for comparison with the results of
\citet{2000ApJ...529.1011P}.  The models of \citet{2000ApJ...529.1011P} did not
account for the temperature distribution across the surface of the
star, so we also calculated a model with $T_\rmscr{eff}=10^6$~K over 
the entire star.

Although some subtle details of the emergent spectra depend on the
value of the surface gravity, we are only concerned with the gross
polarization of the emergent radiation; therefore, we adopt a
canonical value of $g_s=2.4 \times 10^{14}$~cm~s$^{-2}$.  This
simplification reduces the number of atmosphere calculations by a
factor of three.

The polarized spectrum is derived numerically from a self-consistent
solution to the equations radiative transfer for a stationary,
plane-parallel atmosphere in radiative equilibrium.  We adopt the
simplifying assumption that the atmospheric plasma is pure hydrogen in
the limit of complete ionization, for which the opacity sources are
inverse bremsstrahlung, Thomson scattering, and resonant scattering by
protons.  The magnetic field is assumed to be vertically uniform.  The
atmospheric model is obtained by the method of complete linearization
\citep{Miha78}, starting from a power-law prescription for the
conductivity of the plasma as a trial solution \citep{Heyl98rcw103}.
Convergence is achieved $(\delta T/T < 10^{-3})$ typically within
10-20 iterations; flux is conserved to about $10^{-4}$ at all depths.

Radiative transport proceeds in two coupled normal modes of
polarization, uniquely defined for propagation angle $\theta_{B}$ with
respect to the direction of $\vec{B}$.  The magnetic field induces a
strong angular dependence in the plasma opacity, and suppresses the
opacity by a factor $~(E_{\gamma}/E_{cyc}$ in one mode of propagation
\citep{1979PhRvD..19.1684V}.  Consequently, the emergent flux is
dominated by the ``extraordinary mode'' which sees a more transparant
medium over a broad range of $\theta_{B}$.  To the best of our
knowledge, these models are the first models to employ complete
linearization and to consider neutron-star atmospheres where the
magnetic field is slanted with respect to the normal.

\subsection{Integrated Spectrum}
\label{sec:integrated-spectrum}

Once a polarization ``image'' is calculated (see Paper~I for details),
the net polarization seen by an observer is found by integrating the
intensity contributed by each of the normal modes of the atmosphere to
each of the observed Stokes's parameters, adapting the results of
\citet{Pern00axp}, we have
\begin{eqnarray}
S_{i,\nu}&=&\frac{1}{4\pi D^2 (1+z)^3}
\int_0^{b_\rmscr{max}} b d b \int_0^{2\pi} d\beta \times
\nonumber \\
& & ~~~~~~
\left [ 
|\left < S_i | U | X \right >|^2 \;I_{X,\nu (1+z)}(\phi,\beta;\delta,\upsilon) +
|\left < S_i | U | O \right >|^2 \;I_{O,\nu (1+z)}(\phi,\beta;\delta,\upsilon)
\right ]
\label{eq:flux_calc}
\end{eqnarray}
in units of erg cm$^{-2}$ s$^{-1}$ keV$^{-1}$.  Here $x=\sin\delta= b
/ R_\infty$ ($R_\infty=R(1+z)$ and $\delta$ is the apparent zenith
angle of the observer from a point on the surface),
$b_\rmscr{max}=R_\infty$ for $R>3 M$ and $b_\rmscr{max}=3 \sqrt{3} M$
for $R \leq 3 M$.  The gravitational redshift at the surface is given
by $1+z=(1-2M/R)^{-1/2}$. $I_{O,\nu}$ and $I_{X,\nu}$ are the intenisty
per unit frequency per unit solid angle emitted at the surface, in the
ordinary and extraordinary modes respectively.  $\phi$, $\beta$ and $\delta$ 
are defined in \S~2.2 of Paper I.  $\upsilon$ is the azimuth of th
line of the sight from the surface.

The general relativistic effects of light deflection are taken into
account through the ray-tracing function, $\phi(x)$ \citep[][and
\S 2.2 of Paper I]{Page95}). As previously mentioned, if $S_1/S_0=1$ 
is taken to
denote light fully polarized perpendicular to the projection of the
direction of the magnetic dipole moment, symmetry dictates that
$S_2=S_3=0$.  The operator $U$ accounts for the evolution of the
polarization as the photon travels from the surface which Paper~I
discusses in detail.

\section{Results}
\label{sec:results}

\begin{figure}
\plottwo{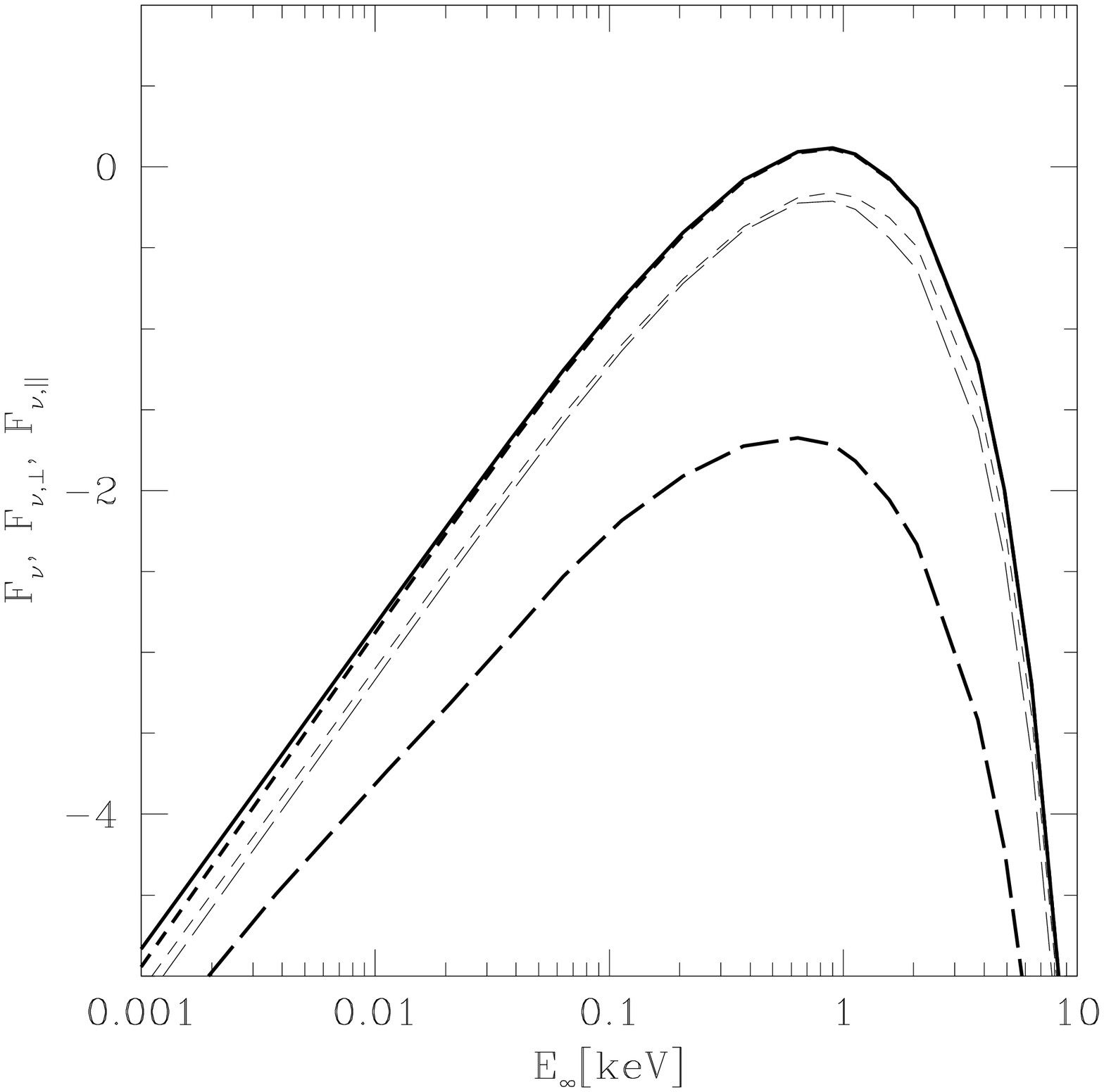}{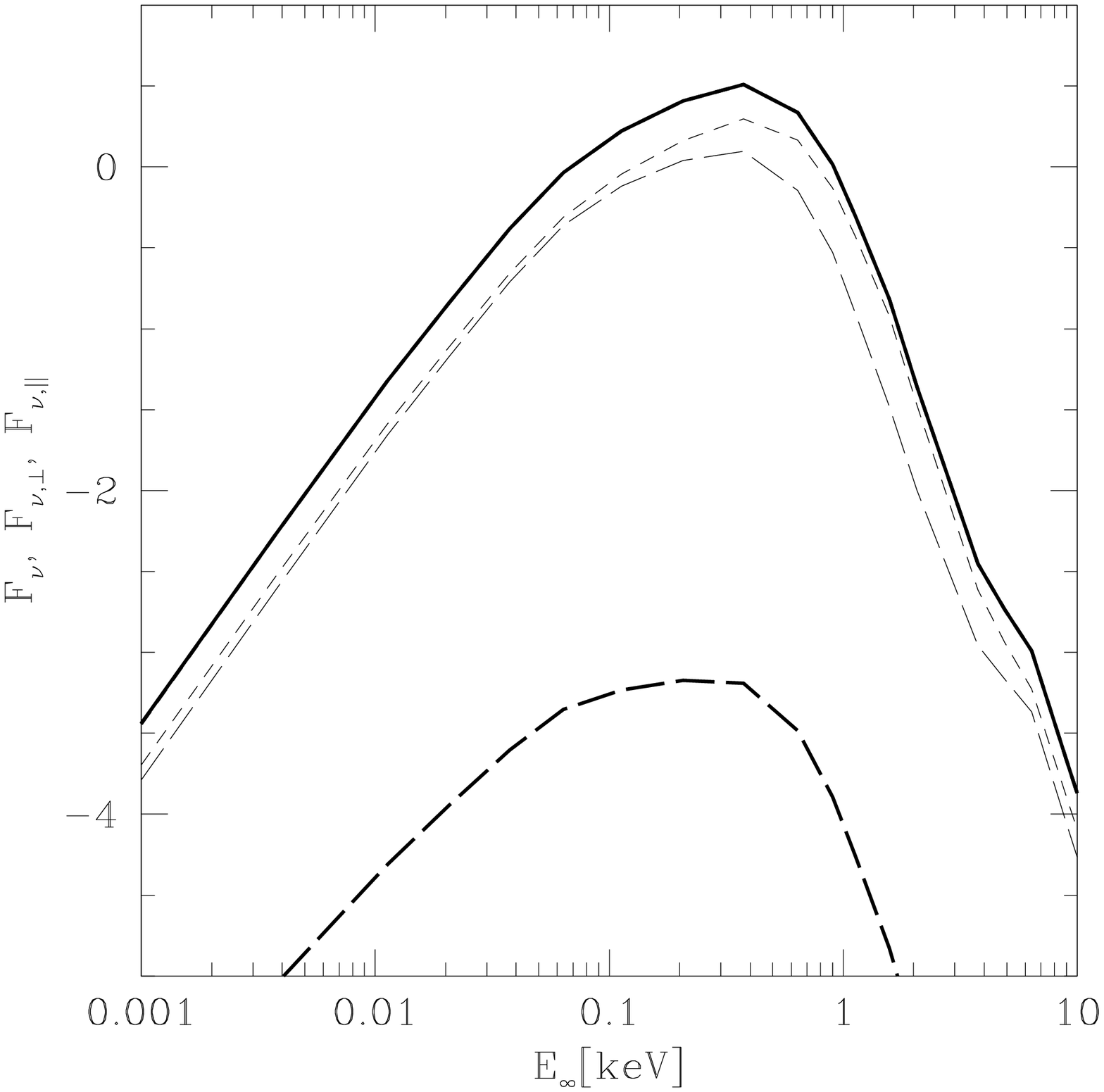}
\caption{The total emergent flux from the visible surface of the star.
The magnetic pole makes an angle of 60$^\circ$ with the line of sight.
The radius of the star is 12~km and its mass is 1.4~M$_\odot$.  The
effective temperature at the magnetic pole is $10^{6.5}$~K. The left
panel is for a magnetic moment of $\mu=10^{30}$~G~cm$^3$ and the right
panel is for $\mu=10^{32}$~G~cm$^3$, corresponding to surface fields
of $\sim 10^{12}$~G and $\sim 10^{14}$~G respectively.  The solid
curve traces the total flux.  The short dashed curve traces the flux
polarized perpendicular to the projection of the magnetic moment in
the plane of the sky.  The long-dashed curve traces the flux polarized
parallel to the projected magnetic moment.  The heavy curves trace the
results including vacuum polarization and the light curves neglect
it.}
\label{fig:spectrumplot}
\end{figure}

By combining the detail calculations of the radiative transfer through
atmosphere with ray tracing through the magnetosphere including the
evolution of the polarization, we obtain spectral energy distrubtions
like those depicted in Fig.~\ref{fig:spectrumplot}.  A comparision of
the total emergent flux in two magnetic fields shows that the more
weakly magnetized atmosphere has a bluer spectrum; \citet{Lloy02} have
noted this effect.  In more detail, if we examine the results
neglecting vacuum polarization (light curves), the emission from the
more weakly magnetized atmosphere is also less polarized.  The most
important trend, which we shall explore in further detail, is apparent
from comparing the light curves with the dark curves which include the
effects of vacuum polarization in the magnetosphere.  In both cases,
the inclusion of vacuum polarization increases the polarized flux from
$\sim 10\%$ to nearly 100\%.

\subsection{Dependence on Stellar Radius}
\label{sec:depend-stell-radi}
\begin{figure}
{ \plottwo{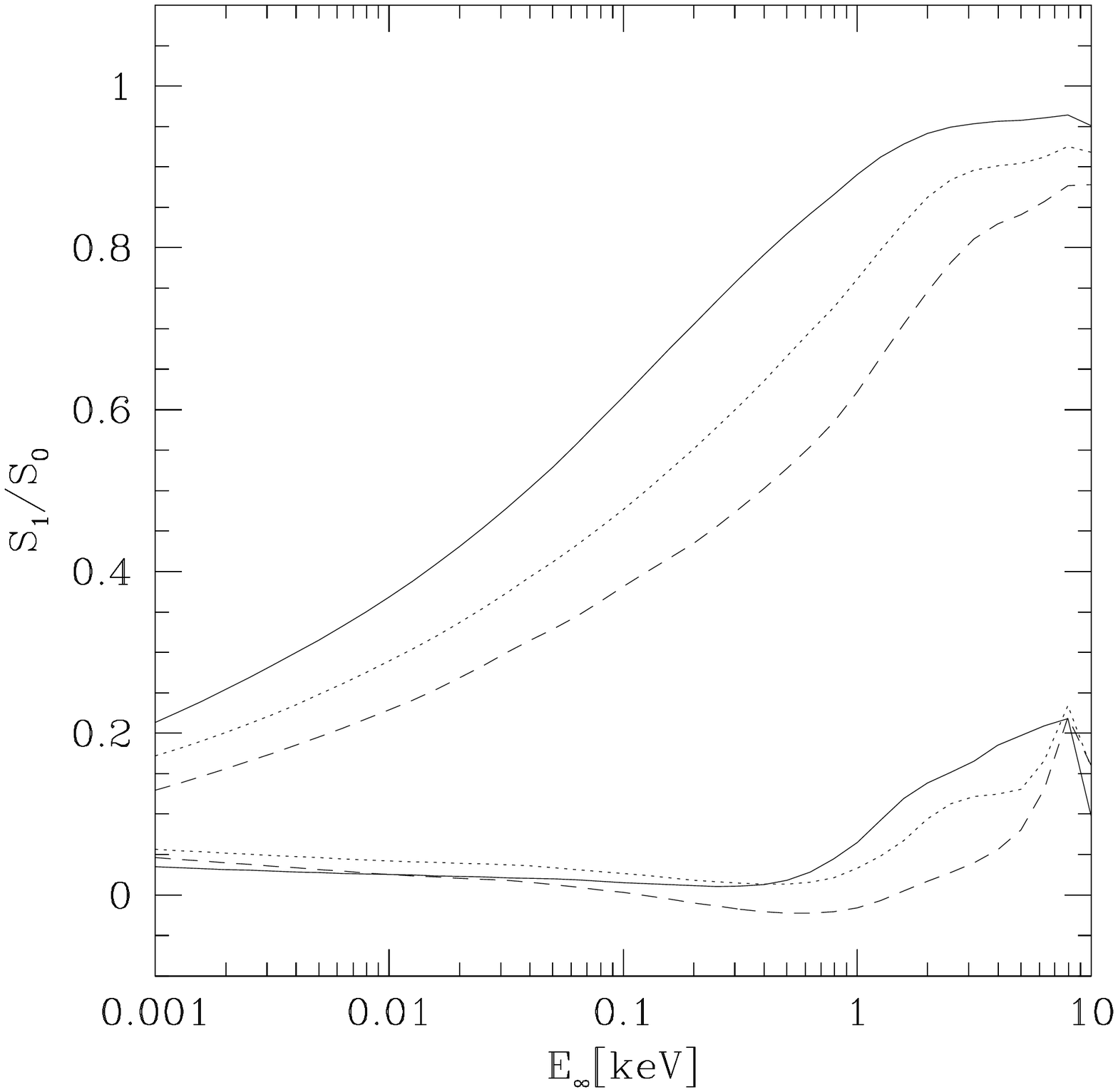}{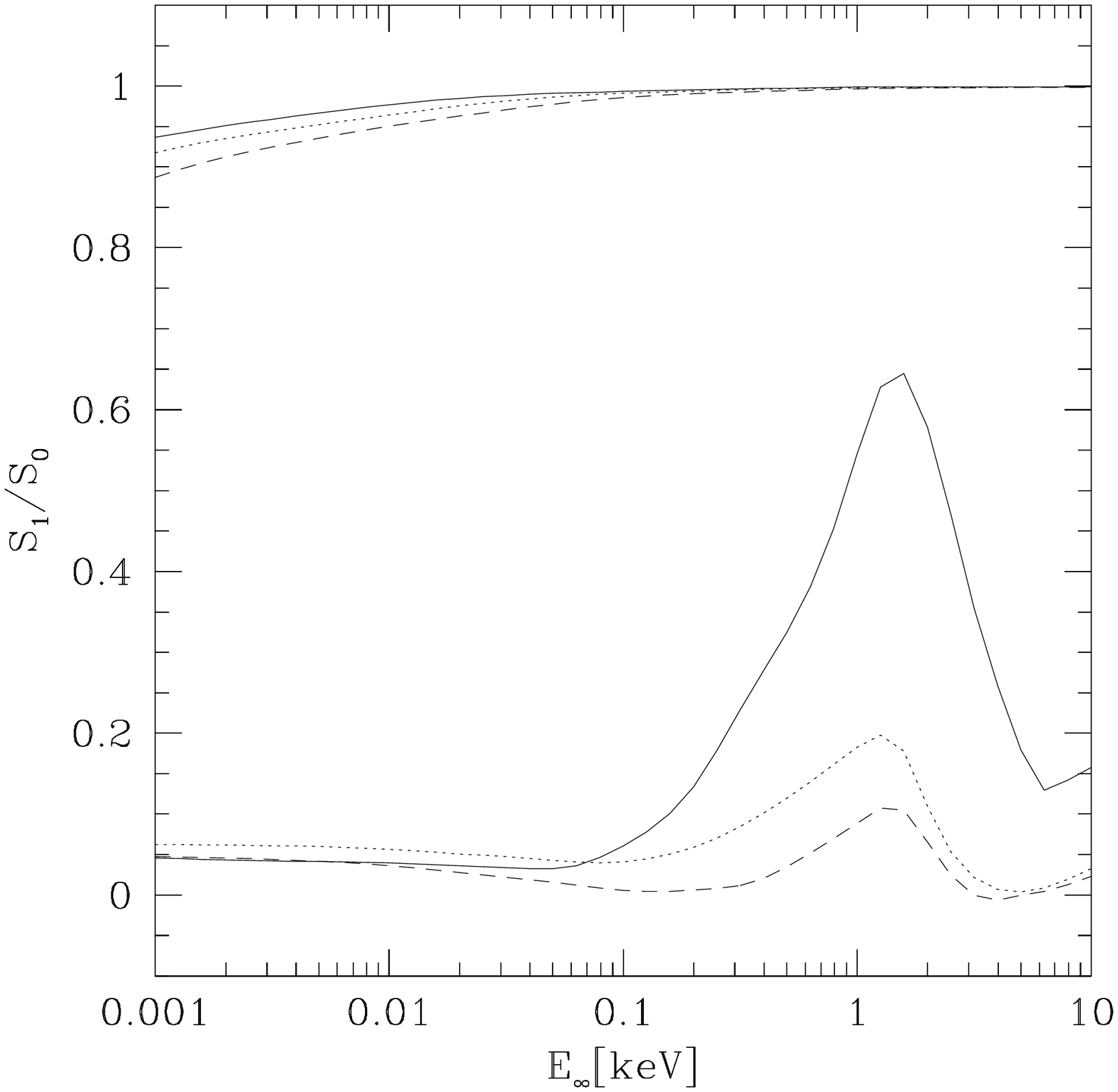} }

{ \plottwo{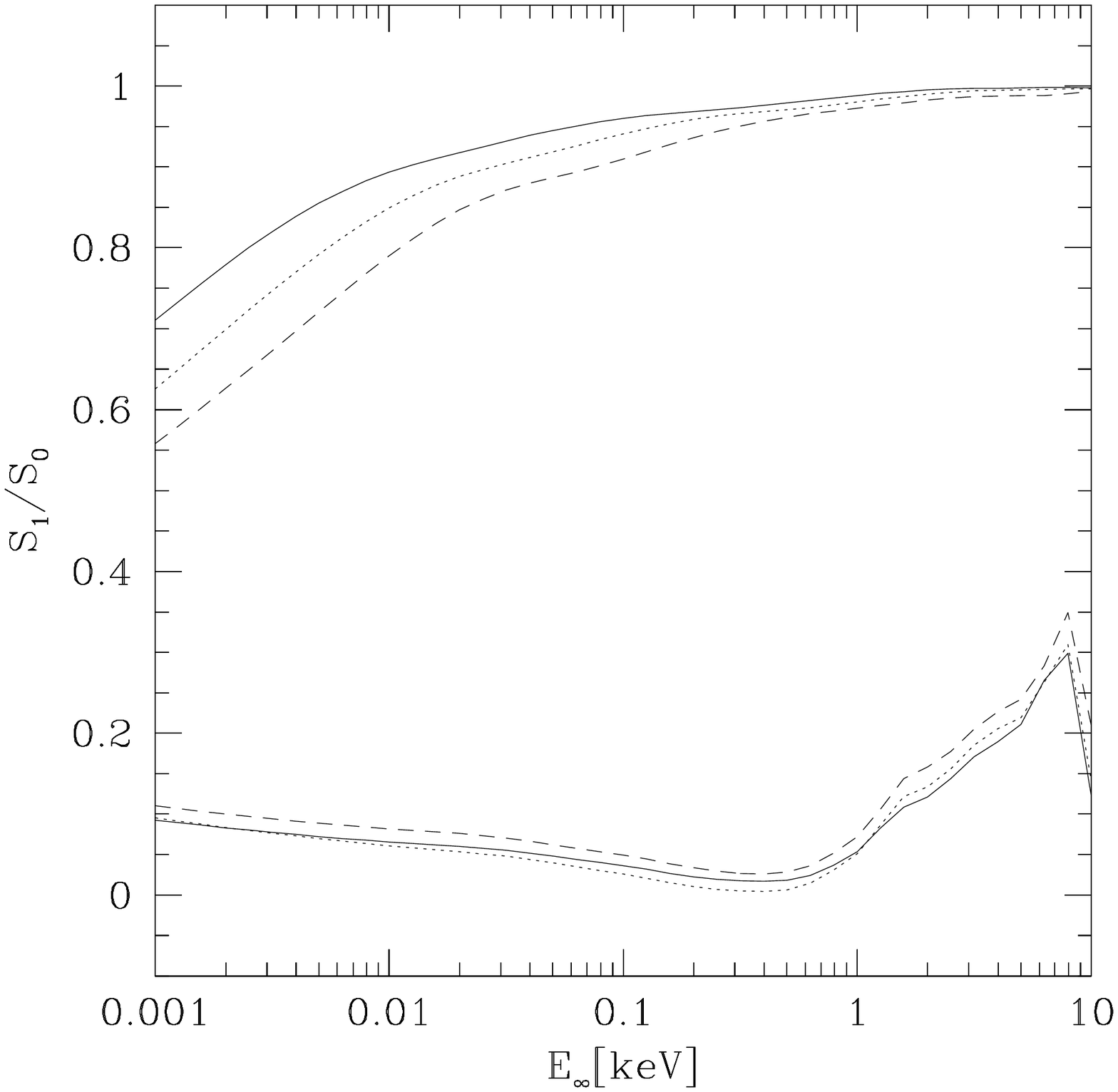}{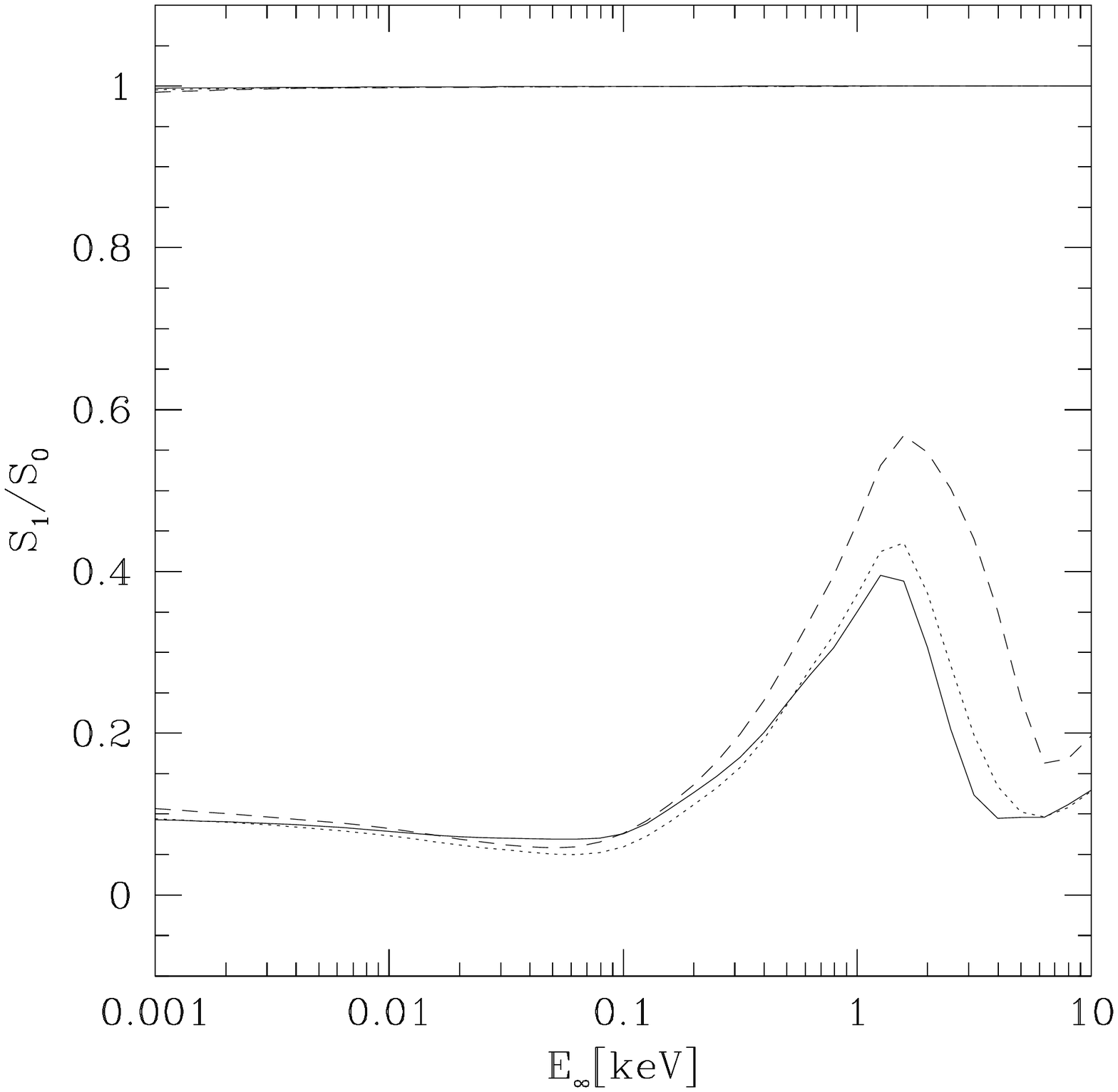}}
\caption{The extent of the polarization as a function of energy, angle
and stellar radius.  $S_1/S_0=1$ denotes light fully polarized
perpendicular to the projection of the dipole into the sky.
$S_1/S_0=-1$ is light fully polarized parallel to this vector.  In the
upper panels, the magnetic pole makes an angle of 30$^\circ$ with the
line of sight.  In the lower panels the angle is 60$^\circ$.
The lower set of curves trace the results without vacuum polarization,
and the results for the upper curves include it.  The solid, dotted
and dashed curves follow the result for 8, 10 and 12~km neutron stars.
The mass of the star is 1.4~M$_\odot$.  The effective temperature at
the magnetic pole is $10^{6.5}$~K. The left panel is for a magnetic
moment of $\mu=10^{30}$~G~cm$^3$ and the right panel is for
$\mu=10^{32}$~G~cm$^3$.}
\label{fig:radiusplot}
\end{figure}

Fig.~\ref{fig:radiusplot} shows the extent of polarization as a
function of energy and radius for $10^{12}$~G and $10^{14}$~G neutron
stars.  The observed net polarization depends strongly on both the
strength of the magnetic dipole moment and the radius of the star.  We
find that smaller stars exhibit higher polarized fractions -- this is
the opposite of the \citet{2000ApJ...529.1011P} result.  Although
gravitational lensing is more important for the more compact stars and
we {\em do} see a larger fraction of the stellar surface, the observed
polarization direction reflects the direction of the magnetic field at
the polarization-limiting radius {\em not} at the surface.  The size
of $r_\rmscr{pl}$ (\cf eq.~16 of Paper I) is independent of the radius
$R$ of the star.  Meanwhile as $R$ decreases the solid angle subtended
at $r_\rmscr{pl}$ by the bundle of rays destined for our telescope
also decreases, thus probing less of the variance in the magnetic
field direction and resulting in a higher polarized fraction.  

Not surprisingly, if we neglect in the birefringence of magnetosphere,
we sometimes find that smaller stars do exhibit smaller polarized
fractions \citep[as][found]{2000ApJ...529.1011P}.  This is most
apparent for the case where the magnetic pole makes an angle of
60$^\circ$ with respect to the line of sight.  If the magnetic pole
makes a smaller angle with the line of sight, the trend of the
polarization with the compactness is less apparent.  Because the flux
from the stellar surface is concentrated near the magnetic poles, only
if the image contains both poles can there be a substantial reduction in
the polarization.  If one pole makes a 60$^\circ$-angle with the line
of sight, the other pole is visible for all three stellar radii that
we have considered and as the stellar radius decreases more of the
second polar region becomes visible, and the total polarization
decreases.  On the other hand, if the pole makes a 30$^\circ$-angle
with the line of sight, only if the stellar radius is less than
8.55~km are both poles visible.  In this case, the trend with radius
is more complicated.  Regardless, because the vacuum birefringence
strongly affects the observed polarization, understanding these trends
in detail only provides a bridge to compare our results with earlier
results with have neglected the QED effects.

\subsection{Angular Dependence}
\label{sec:angular-dependence}

Comparing the results depicted in the upper and lower panels of
Fig.~\ref{fig:radiusplot} explores the extent of polarization as a
function of frequency and angle.  The results here confirm the results
of \citet{Heyl01polar} (see their Fig.~5).  The extent of the observed
polarization increases with the angle between the line of sight and
the dipole axis.  This effect is apparent both with and without vacuum
polarization in the magnetosphere.  The source of this trend is quite
straightforward.  If the observer is looking straight down onto the
magnetic pole of the star, the net polarization must vanish by
symmetry, independent of vacuum birefringence.  As this angle
increases so to must the net polarization.  Whether the net
polarization attains a maximum before the angle reaches ninety degrees
will depend on the details.  If polarization-limiting radius is much
larger than the stellar radius, the polarization will increase until
the angle between the line of sight and the pole reaches ninety
degrees.  This is also the case if one neglects the gravitational
defocussing of the stellar surface, \ie in the limit of $R \gg G
M/c^2$.  If $R \sim G M/c^2$ and the polarization-limiting radius is
not much larger than $R$, the behavior may be more complicated.

\subsection{Comparison with Earlier Results}
\label{sec:comp-with-earl}

These results presented here agree well with the trends explored in
\citet{Heyl01polar}.  Specifically, the polarization of the flux
increases with the frequency of the radiation, with the angle between
the magnetic dipole moment and the line of sight and with the strength
of the magnetic dipole moment, and decreases as the radius of the star
increases.  All of these effects are driven by the QED-induced vacuum
polarization in the magnetosphere.   The most recent comparable work 
other than our own is that of \citet{2000ApJ...529.1011P}.  
The study of \citet{2000ApJ...529.1011P} differs from ours in three important 
respects.  First, \citet{2000ApJ...529.1011P} neglected the effects of vacuum 
polarization in the magnetosphere.  Second, \citet{2000ApJ...529.1011P} did not 
account for the fact that the temperature across the surface of the
neutron star varies.  Third, \citet{2000ApJ...529.1011P} used the
diffusion approximation in calculating their atmosphere models while
we used a more accurate complete linearization treatment.  

\begin{figure}
\plottwo{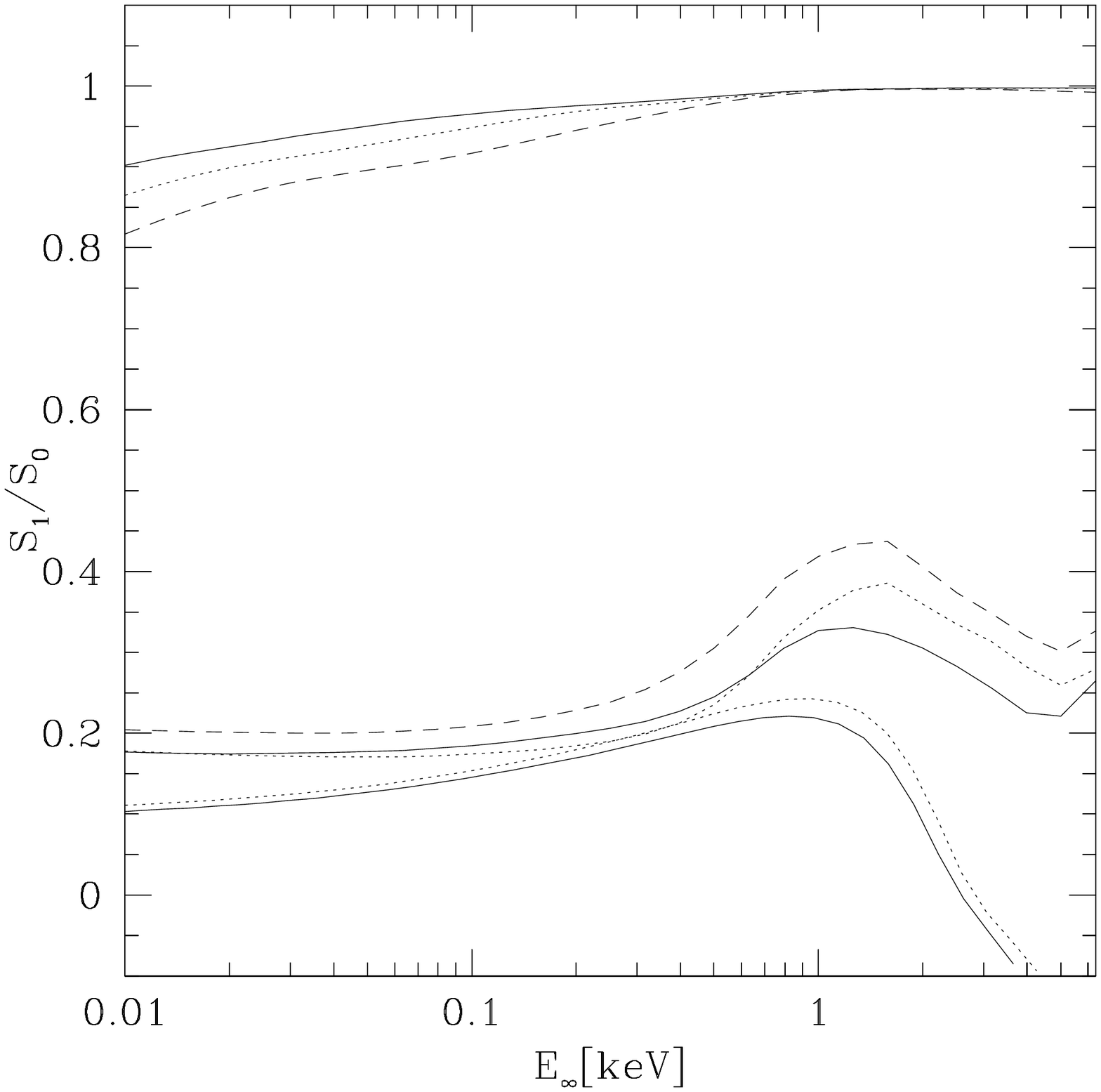}{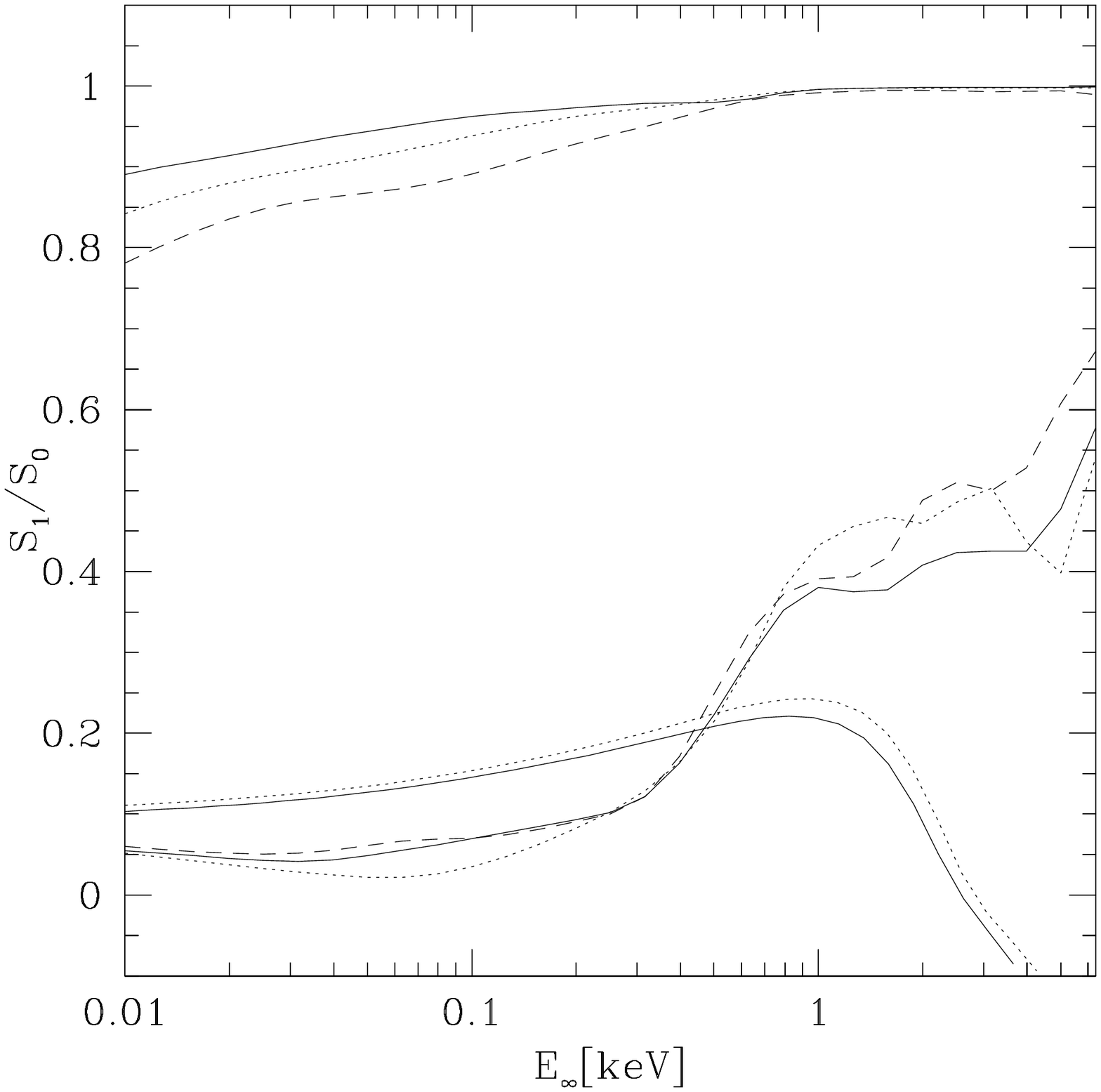}
\caption{Comparison with the results of \citet{2000ApJ...529.1011P}.
  The solid line, dotted and dashed curves trace results for 8, 10,
  and 12 km neutron stars with $M=1.4$M$_\odot$ and
  $\mu=10^{30}$~G~cm$^3$.. In both panels, the magnetic pole makes a
  60$^\circ$ with the line of sight.  The left panel assumes that the
  temperature of the star is constant over the surface as
  \citet{2000ApJ...529.1011P} did.  The right panel accounts for the
  variable temperature of the surface.  The uppermost curves include
  the radiative corrections of QED.  The middle curves are our resutls
  without radiative corrections and the lowermost curves are the
  results obtained by \citet{2000ApJ...529.1011P}.}
\label{fig:pavlcomp}
\end{figure}

By shutting off the vacuum polarization and the variation of the
temperature across the surface we can attempt to reproduce the results
of \citet{2000ApJ...529.1011P}.  This is depicted in the left panel of
Fig.~\ref{fig:pavlcomp}.  The lower two curves are the results of
\citet{2000ApJ...529.1011P} at $T_\rmscr{eff}=10^6$~K for the two
smaller radii depicted in Fig.~\ref{fig:radiusplot}.  The middle three
curves follow the results neglecting vacuum polarization for an
isothermal surface at the same effective temperature.  The qualitative
agreement is good but quantitative disagreements arise due to the
differing treatment of radiative transfer.  The upper three lines
include vacuum polarization for these same parameters.

The right panel of Fig.~\ref{fig:pavlcomp} depicts models in which the
surface temperature varies according to
Eq.~\ref{eq:temp-distribution}.  In this case a small portion of the
star contributes the bulk of the flux at high energies, so not
surprisingly the extent of polarization is larger above a few hundred
electron volts than in the model depicted in left panel.  The extent
of the polarization at low energies is somewhat lower reflecting differences in 
the detailed treatment of the atmosphere upon which the low-energy flux
strongly depends.

\section{Discussion}
\label{sec:discussion}

We have found that the treatment of the vacuum polarization of the
magnetosphere surrounding a neutron star is crucial to determine the
polarized emission from the surface accurately.

We have also shown the effect that real atmospheres have on the
polarization.  The net polarization observed in this case is somewhat
reduced since a surface element will have a lower intrinsic
polarization to begin with, such that even complete alignment will not
result with $100$\% polarization. Anomalous X-ray Pulsars (AXPs), if
they are magnetars (neutron stars with surface magnetic fields of
$10^{14}$ G or larger), should exhibit an extremely large polarization
even at optical wavelengths.  This could potentially be detected in
the observed optical counterpart to an AXP, thus possibly verifying
that AXPs are indeed magnetars.

In the optical/UV the polarized fraction depends quite strongly on
$\mu$ and $R$.  In \figref{radiusplot}, the magnetic dipole moment
varies by a factor of one hundred between the left and right panels.
On each panel of curves, the radius varies between eight and twelve
kilometers.  Decreasing the stellar radius by a factor of $3/2$
increases the observed polarized fraction by a factor of 1.3 to 1.7
for the more weakly magnetized stars, similar to that achieved by
increasing the dipole moment by a factor of six, because the polarized
fraction is approximately a function of $b_\rmscr{max}/r_\rmscr{pl}$.
If the magnetic dipole moment of a neutron star is moderately
constrained (\eg by spin down), polarization observations can use this
strong dependence to constrain the stellar radius.

Without the effects of QED alignment, the highest expected
time-resolved polarized fraction is about 15\%-25\% under the most
favorable conditions.  However, the time-averaged measurement will
typically be smaller.  With the effects of QED alignment taken into
account, the expected time-resolved polarization at X-rays is close
to the maximum emitted by the atmosphere (50\%-95\%), with the
time-averaged case being somewhat lower, but still larger on average
than the highest time-resolved net polarization expected if QED is
neglected.  Under coincidentally unfavorable conditions, the
time-averaged case can nevertheless be averaged to a low value.
Namely, without time-resolved polarimetry one can prove, but not
necessarily disprove the importance of QED polarization alignment. In
such a case, it would be advantageous to look at the thermal emission
of several NSs. If time-resolved polarimetry is available, then even
one object could be used to verify the birefringence of the
magnetized vacuum due to QED.
 
\acknowledgments
Support for this work was provided by the National Aeronautics and
Space Administration through Chandra Postdoctoral Fellowship Award
Number PF0-10015 issued by the Chandra X-ray Observatory Center, which
is operated by the Smithsonian Astrophysical Observatory for and on
behalf of NASA under contract NAS8-39073. 

\newcommand{\mn}{MNRAS}

\bibliographystyle{apj}
\bibliography{mine,physics,ns,gr,polarpaps}

\end{document}